# A systematic review of safety-critical scenarios between automated vehicles and vulnerable road users

Aditya Deshmukh[1], Zifei Wang[2], Aaron Gunn[3], Huizhong Guo[4], Rini Sherony[5], Fred Feng[6], Brian Lin[7], Shan Bao[8], Feng Zhou[9]

[1,2,6,8,9]University of Michigan Dearborn, Dearborn, MI
[3]University of Michigan, Ann Arbor, MI
[4,7,8]University of Michigan Transportation Research Institute, Ann Arbor, MI
[5]Toyota Motor North America, Ann Arbor, MI

Automated vehicles (AVs) are of great potential in reducing crashes on the road. However, it is still complicated to eliminate all the possible accidents, especially those with vulnerable road users (VRUs), who are among the greater risk than vehicle occupants in traffic accidents. Thus, in this paper, we conducted a systematic review of safety-critical scenarios between AVs and VRUs. We identified 39 papers in the literature and typical safety-critical scenarios between AVs and VRUs. They were further divided into three categories, including human factors, environmental factors, and vehicle factors. We then discussed the development, challenges, and possible solutions for each category. In order to further improve the safety of VRUs when interacting with AVs, multiple stakeholders should work together to 1) improve AI and sensor technologies and vehicle automation, 2) redesign the current transportation infrastructure, 3) design effective communication technologies and interfaces between vehicles and between vehicles and VRUs, and 4) design effective simulation and testing methods to support and evaluate both infrastructure and technologies.

## INTRODUCTION

Vulnerable road users (VRUs) (e.g., pedestrians, cyclists, skate-boarders, e-scooter) are among the greater risk than vehicle occupants in traffic accidents. VRUs accounted for 26% of all road traffic death globally (Sun et al., 2022). In the United States, pedestrian fatalities increased by 10.63% and 9.63% in 2015 and 2016 while pedal-cyclist fatalities increased by 30% between 2009 and 2018 (NHTSA, 2020). Although automated vehicles (AVs) are expected to improve the safety, mobility, and efficiency of the transportation system (Ayoub, Zhou, Bao, & Yang, 2019) there are still challenges when interacting with VRUs in mixed traffic in the near future. Due to the complexity of various of interaction scenarios between AVs and VRUs, it is important to conduct a systematic review of the safety-critical scenarios in the literature.

Therefore, in this study, we 1) conducted a systematic review to understand the major factors that lead to possible accidents between AVs and VRUs, 2) proposed a taxonomy of the major factors involved in these safety-critical scenarios, and 3) discussed the possible solutions to handle these scenarios.

## METHOD

We conducted a systematic search of related studies by following the PRISMA (Preferred Reporting Items for Systematic Reviews and Meta-Analyses) process (Moher, Liberati, Tetzlaff, Altman, & Group, 2009) in different databases, including ACM digital library, IEEE Xplore, ScienceDirect, and IIHS (Insurance Institute for Highway Safety). The sets of domain-specific searching key terms are:{"automated vehicle", "automated driving", "autonomous vehicle", "ADAS", "ADS", "bicyclists", "corner cases", "crashes", "cyclists", "pedestrians", "VRU", "E-scooter", Micro mobility", "skateboard"}. We used a combination of one or more key terms for searching. Key terms were searched within metadata (i.e., title, abstract, and keywords). Papers were restricted to journal publications, conference proceedings, and theses. Eligible studies were within a period from January 2010 to January 2022.

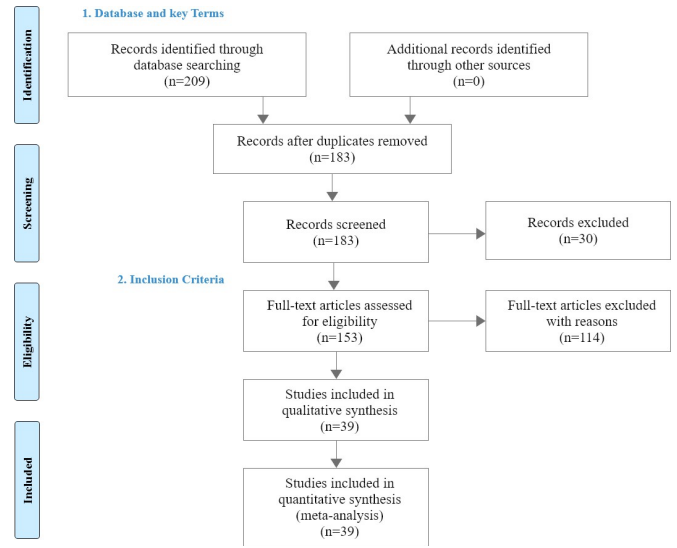

*Figure 1*. PRISMA overview method for literature searching.

In total, 209 papers were identified in the selected databases. We identified 183 unique papers in total by removing 26 duplicate papers with the following eligibility criteria: 1) An empirical study (e.g., crash data analysis, human-subject experiment, or on-road test/observation), or 2) at least one incident or safety-critical scenario related to AV-VRU had to be reported. After scanning all 183 papers for eligibility based on abstract and methodology, we identified 39 relevant papers for further assessment and systematic review per the above criteria. The paper selection process is shown in Figure 1.

## RESULTS

We identified 47 scenarios involving automated vehicles and VRUs from the selected papers. The identified scenarios were grouped into three main categories, including human factors (44.7%), environmental factors (27.7%), and vehicle fac-tors (27.7%). Within each main category, subcategories were also defined to further explain the scenarios. The details are shown in Figure 2. For the scenarios related to human factors, most of them were about VRU behavior (76.2%), followed by driver's behavior (14.3%), and other driver's behavior (9.5%). For the scenarios related to environmental factors, the majority were about occlusion (76.9%), followed by road conditions (15.4%) and light condition (7.7%). For the scenarios related to vehicle factors, most of them were about system limitations (46.2%) and system failures (38.5%), followed by algorithmic decision-making (15.4%).

Table 1. *Typical scenarios related to human factors*

| Type | Description |
|---|---|
| Bicyclists | • Riding alternatively from one side of the lane to the other<br>• Standing up to pedal to accelerate<br>• Riding in the opposite direction of the traffic<br>• Cross an intersection with vehicles around<br>• A bicyclist initiating a left-turn/U-turn maneuver at a signalized intersection and having to use a full travel lane while preparing to make a left turn |
| Pedestrians | • Crossing a nonprivileged road<br>• Not yielding to the incoming traffic<br>• Crossing the road at the intersection with the red light on<br>• Pedestrians standing on the road for an indefinite period<br>• Pedestrian leaving the sidewalk in such a short distance so that the vehicle could not stop<br>• A distracted pedestrian crossing a signalized intersection when the conflicting driver had the right-of-way<br>• Children crossing the road<br>• A pedestrian jaywalking while crossing the road<br>• A pedestrian waiting in the middle of the road to let the vehicle cross |
| Other drivers | • Sudden lane changing of the front vehicle<br>• Sudden cuts of the other vehicles<br>• Wrong way driving of the motorist or other vehicles |
| Drivers | • AV crashing into a bus while leaving the parking space as it assumed that the bus driver would give the right of way |

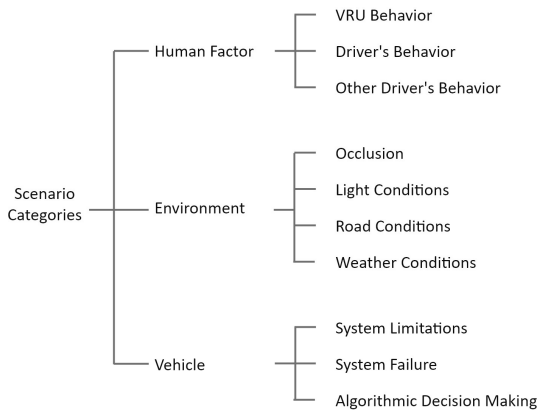

*Figure 2*. Classification of Scenarios

## Human Factors

The identified scenarios related to human factors are shown in Table 1 and they are divided into five different types, including e-scooters, bicyclists, pedestrians, other drivers and drivers of the AV. The majority of the scenarios are associated with pedestrians.

*Challenges.* The major challenges we identified related to human factors were summarized as follows:

(a) Unpredictability of VRU behaviors
(b) Unpredictability of other vehicles
(c) Understand VRU intentions
(d) Complexities of various safety-critical scenarios
(e) Incompatibility/inadequacy of current AV infrastructure
(f) Mixed traffic between AV and conventional vehicles
(g) VRUs occlusion by objects
(h) Inadequacy of testing and evaluation methods
(i) Density of VRUs in complicated scenarios
(j) Lack of effective communication between VRUs and AVs

*Proposed Solutions.* While these scenarios create challenges for AVs, various solutions were proposed in the literature as summarized below.

**Machine Learning Methods:** Many researchers attempted to deal with interactions with VRUs with machine learning models to understand their behaviors and intentions so that AVs can take proactive responses to avoid potential accidents. In order to improve cyclist safety, Gu and Zhou (2017) proposed a solution with smartphones, which used sensors and GPS data to train a machine learning model to monitor the bicyclist's behavior to improve safety. The results showed that the proposed solutions achieved accuracy between 87% and 90% for detecting the three given scenarios. Garate and Bours (2012) proposed to detect VRUs using machine learning models, based on which an automatic braking system was used to reduce the effect of collision. Such pre-collision systems might create false alarms if the intentions of the VRUs are not well understood, which is one of the main challenges. Liu et al. (2016) analyzed the behavior of bicyclists by measuring the time to collision, lateral positioning, and the speed to inform the pre-collision system to interpret their intentions.

**Redesign Transportation Infrastructure:** The existing transportation infrastructure seems to be inadequate to address various scenarios between AVs and VRUs in mixed traffic. For example, Kiss and Berecz (2019) discussed the safety issues related to AVs in different scenarios and suggested that AV systems should be designed to be aware of the various situations and problems of infrastructure, such as road layouts, vehicle systems, and pedestrian behaviors. Mako (2015) investigated pedestrian accidents at crossings and found that safety measures, such as refuge islands, traffic signals, and flashing lights improved driver behavior. Polnnaluri and Alluri (2021) proposed using intelligent transportation systems to address issues, such as distracted pedestrians crossing when drivers have the right-of-way, bicyclists making left-turns or U-turns, and wrong-way driving by motorists. However, these technologies are not yet widely adopted.

While comparing to road infrastructure, network systems are promising to improve the safety of the VRUs. Sam et al. (2015) identified situations where accidents occurred due to obstructed views of VRUs caused by buildings, junctions, or traffic. They proposed a solution by creating an alert system based on a vehicular Ad Hoc network to give drivers advanced warning and increase their reaction time. However, testing such a system in real-world scenarios could pose risks. To overcome this limitation, Heinovski et al. (2019) developed a virtual model to simulate interactions between vehicles and cyclists at intersections and investigate and record critical scenarios of cyclists' behavior. Moreover, recent advancements in communication technology, such as Wifi Direct and Dedicated Short-Range Communication enabled it to gain attention for vehicle-pedestrian communication. Tahmasbi-Sarvestani et al. (2017) proposed a vehicle-to-pedestrian safety system to extend situation awareness and hazard detection capabilities using SAE J2735 Personal Safety Message (PSM). The system was implemented and tested in intersection crossings between vehicles and pedestrians. They found such a communication system provided promising results to improve VRU safety.

**Investigating VRU Behaviors:** To increase the traffic safety of VRUs, the behavior of pedestrians and cyclists should be taken into consideration while designing the safety systems, such as trust, willingness and culture that affect the behaviour of the people. Wong (2019) examined the factors affecting people's behavior and emotions toward AVs and found that those presented with subjective information about an AV crash scenario showed less trust in AVs compared to those with factual information. Sikkenk and Terken (2015) surveyed people's willingness to display polite traffic behavior when interacting with AVs and found that the willingness to give the right of way was influenced by various factors, such as weather, the vulnerability of the road users, and driving styles of the participants. Wang et al. (2016) compared traffic safety cultures between Sweden and China by testing pre-designed incidents similar to the real traffic situations (e.g., sudden lane changing or cuts of the front vehicle, pedestrian suddenly coming to the road). They found that the designed advisory traffic information system increased the safety of the drivers in both countries. Moreover, increasing the traffic safety for children is important because of children's inability to scan the environment while crossing the road. Riaz and Cuenen (2019) developed a platform to teach children traffic safety rules by including four modules based on real-life situations. They found that children performed better in the familiar situations compared to unfamiliar situations.

**Simulation and Testing:** Researchers aim to increase safety for pedestrians and drivers using AV systems, but real-life testing is difficult and risky due to unpredictable human behavior and the challenge of gathering data on VRUs. Jannat et al. (2020) developed a pedestrian technology test bed and evaluated three different systems (camera-based, camera-radar fusion, and smartphone-based P2I) in four test scenarios, finding that the camera-based systems were vulnerable to environmental conditions while the smartphone-based P2I applications were not. Doric et al. (2016) introduced and evaluated a simulation-based system to mix human behavior when testing traffic safety systems, achieving an acceptable level of realism in a pedestrian-failed-to-yield scenario. AVs are improving traffic safety by attempting to eliminate human errors. However, there is still a need to understand the implications of AVs on traffic safety and researchers used different testing methods to assess the safety benefits. Kutela et al. (2022) analyzed AV crash data to explore the involvement of VRUs in traffic and found that bicyclists and cyclists were more likely to be involved in AV crashes and that crosswalks, intersections, and traffic signals were key factors. Hollander et al. (2020) developed a game to gather data on pedestrian behavior, finding that the game's scenarios correlated with real-world patterns and could be used to create a system for vehicle-pedestrian safety.

Table 2. *Typical Scenarios related to environmental factors*

| Type | Description |
|---|---|
| Occlusions | • Pedestrian hiding by the corner of the building while walking, not leaving enough time for the car to stop<br>• A pedestrian walking along/against the traffic and obscured by the other traffic<br>• Two pedestrians start crossing the street right in front of the driver. Of these two, one starts walking first and is partly hidden in the pillar blind spot.<br>• A car taking a left turn at the intersection and the VRU is obscured by the opposing vehicle<br>• A pile of snow in the sidewalk obscures the pedestrian's view and the driver's view.<br>• Pedestrians being blocked by the corner of buildings |
| Light Conditions | • Cyclists failing to detect right-turning vehicles under poor lighting conditions |
| Road Conditions | • Two vehicles parking perpendicularly on both sides of the residential road creating a bottleneck<br>• Layout of the road is not proper making it difficult to distinguish the main road from the residential road<br>• Excessively worn road markings<br>• Previous road markings are visible or have been adjusted to road construction |

**Environmental Factors**

Table 2 shows the scenarios related to environmental factors in three different types, including occlusion, light conditions, and road conditions. The complexity of those scenarios are the main challenges facing AVs.

*Proposed Solutions.* We summarized the solutions below:
**Communicating with VRUs:** The current communications between vehicles can be extended to those between vehicles and VRUs for improved traffic safety. Segata et al. (2017) developed a probability framework to inform drivers to prevent collisions at intersections where a vehicle turning left might collide with a bicyclist. External human-machine interfaces (eHMI) on AVs are another hot research topic for communicating with VRUs. Rettenmaier et al. (2020) evaluated different eHMIs by in a driving simulator, where the AV and other human drivers communicated at a road bottleneck. The result showed that the selected e HMIwas able to significantly reduce the passing time and crashes with human drivers. However, communicating in the high traffic area is still a challenge. Rostami et al (2016) proposed to evaluate the performance and channel load for the pedestrian to vehicle transmission in a high-density pedestrian scenario. Their results showed that the performance requirements were hard to meet.

**Investigating Environmental Factors:** Advancements perception technology is increasing AV performance. However, it is still a major issue for scenarios where objects are hidden or cannot be observed directly. Hoermann et al. (2017) proposed a safe approach for vehicles at intersections where pedestrians are hidden by buildings or vegetation, using free sensor fusion to calculate hidden regions and safely navigate turns. Chen et al. (2016) showed that the main factors affecting the brake response time were visibility in the darkness, which shortened the time to brake, intersections, and the number of potential threat vehicles. Sherony and Zhang (2015) analyzed the pre-crash scenarios and crash analysis between VRUs and vehicles and found that drivers of age between 25-30 years were seen to hit the pedestrian and bicyclists when crossing the road due to road conditions and lighting conditions.

Table 3. *Typical Scenarios related to vehicle factors*

| Type | Description |
|---|---|
| System Limitations | • Short takeover lead time while in the heavy traffic density<br>• Digitally displayed imagery is perceived as a real object<br>• AV is not able to operate in adverse conditions and without visible lane markings<br>• Interaction of people of different cultures with AVs<br>• Visible light communication system is ineffective in dusk and radar is ineffective when pedestrians are stationary |
| System Failure | • The AV shows wrong contradicting information on the external display to the VRUs<br>• The ADAS cannot detect the imagery that appears briefly<br>• Sensors failed to detect roadway infrastructure<br>• A vehicle sends false information to another vehicle causing the other vehicle to act and leading to fatal collision |
| Algorithmic Decision Making | • A situation where car has to make a decision. If the AV continues ahead it will hit and kill a group of pedestrians, including three adults and a dog, crossing on a red light. If the AV swerves, it will hit a barrier and kill its passenger |

**Vehicle Factors**

The identified scenarios related to vehicle factors are shown in Table 3 in three different types, including system limitation, system failure, and algorithmic decision making.

*Challenges.* We identified the major challenges described as 1) limitations of current sensor and perception technologies, especially in adverse conditions, 2) incompatibility with current infrastructure, 3) regulation, cultural differences across countries, and 4) security issues of the communication technology.

*Proposed Solutions:* We discuss the solutions proposed in the literature as summarized below:

**Sensor Technology:** AVs are heavily dependent on sensor technologies to increase traffic safety and thus advancements in such technologies can reduce VRU fatalities. Combs et al. (2019) examined the potential of AVs to increase traffic safety by analyzing nearly 5000 pedestrian-related incidents, finding that combining visible-light cameras, light detection and ranging, and radar sensors was more effective than a single sensor in reducing incidents. Recent studies suggest that improving road infrastructure compatibility with AV sensors is important for improving traffic safety. Chipengo and Commens (2019) found that the high radar cross-section of current road guardrails and construction steel plates could trigger false alarms and ignore important signs. They demonstrated two techniques to reduce radar cross-section and improved the effectiveness of sensors used in ADAS systems for vehicle functions like steering, braking, and obstacle alerts. But one challenge of the ADAS system is that they might be unable to differentiate the real obstacles/object from the phantom objects. Nassi et al. (2020) demonstrated a scenario in which the ADAS systems, such as Mobileye 630 and Tesla Model X, recognized phantom depth-less images that appeared for split seconds as a real objects. They also proposed a countermeasure model to determine the authenticity of the object with the help of camera sensors.

Another challenge of the sensor technology is its reliability in adverse conditions for SAE Level 3 and above AVs. For example, Du et al. (2020) found that in the scenarios of the high cognitive load and high incoming traffic density, drivers had the worst performance and lower taker-over readiness. Utriainen (2020) studied the potential impact of AVs on pedestrian safety in adverse weather conditions and evaluated the safety impacts in the scenarios based on the AV's ability to operate in snowy and low-light conditions. The result showed that 28 percent and 73 of fatal crashes could have been avoided by Level 4 and 5 AVs, respectively.

**False or Miscommunication:** Traffic safety can be compromised by vehicles sending malicious and false information to other vehicles in vehicle-to-vehicle communication. Sarker and Shen (2018) studied a scenario where an AV system sent false information to another vehicle which caused a safety-related incident. Thus, it is important to detect such malicious information. The miscommunication using eHMIs between AVs and VRUs can also lead to safety issues. Hollander et al. (2019) investigated the trust of pedestrians in a scenario where the external displays showed misinformation or contradicting information at the intersection. The results showed that the misinformation decreased the trust of pedestrians and or worsened the overtrust even in the case of malfunctioning of the vehicle. Further research in the domain of vehicle-to-VRU communication is needed to increase the safety of VRUs.

**Driver Assistance Technology:** Issues with other assistance technologies, such as navigation, crash avoidance, and automation level also played a role in accidents between VRUs and AVs. For example, Lin et al. (2017) identified the key reasons for accidents caused by navigation systems, such as missing road characteristics data, weather conditions, poor audio and visual instructions from 158 news reports. Good et al. (2017) analyzed how well a crash avoidance system performed in bicycle crash scenarios. Their findings indicated that the crash avoidance system did not offer significant safety or crash avoidance benefits. As a result, further work is required to enhance the effectiveness of crash avoidance systems. Wotton et al. (2022) studied who was responsible for a crash involving an AV that hit a pedestrian due to system failure while the driver was distracted. They looked at four levels of vehicle automation and found that despite differences in drivers' behaviors, the drivers were deemed responsible for the accident even though their behavior did not have a significant impact on the outcome.

## CONCLUSIONS

In this paper, we conducted a systematic review of safety-critical scenarios between AVs and VRUs. We identified three major categories, including human factors, environmental factors, and vehicle factors. Within each category, we identified detailed scenarios from the literature and discussed the proposed solutions that showed the advancement in improving safety of VRUs in interacting with AVs. However, due to the complexities of various scenarios, more work is needed to address the challenges, including but not limited to 1) understanding intentions and behaviors of VRUs, 2) improving the current transportation infrastructure, 3) communicating with VRUs effectively, 4) improving the sensor technologies in AVs, 5) building and utilizing reliable simulation and testing methods, and so on.

## ACKNOWLEDGEMENT

This work was supported by Toyota Motor North America.

## REFERENCES

Ayoub, J., Zhou, F., Bao, S., & Yang, X. J. (2019). From manual driving to automated driving: A review of 10 years of autoui. In *Proceedings of the 11th international conference on automotive user interfaces and interactive vehicular applications* (pp. 70–90). New York, NY, USA: ACM.

Chen, M., Zhu, X., Ma, Z., Li, L., Wang, D., & Liu, J. (2016). Brake response time under near-crash cases with cyclist. In *2016 ieee intelligent vehicles symposium (iv)* (pp. 80–85).

Chipengo, U., & Commens, M. (2019). A 77 ghz simulation study of roadway infrastructure radar signatures for smart roads. In *2019 16th european radar conference (eurad)* (pp. 137–140).

Combs, T. S., Sandt, L. S., Clamann, M. P., & McDonald, N. C. (2019). Automated vehicles and pedestrian safety: exploring the promise and limits of pedestrian detection. *American journal of preventive medicine*, *56*(1), 1–7.

Doric, I., Frison, A.-K., Wintersberger, P., Riener, A., Wittmann, S., Zimmermann, M., & Brandmeier, T. (2016). A novel approach for researching crossing behavior and risk acceptance: The pedestrian simulator. In *Adjunct proceedings of the 8th international conference on automotive user interfaces and interactive vehicular applications* (pp. 39–44).

Du, N., Kim, J., Zhou, F., Pulver, E., Tilbury, D. M., Robert, L. P., ... Yang, X. J. (2020). Evaluating effects of cognitive load, takeover request lead time, and traffic density on drivers' takeover performance in conditionally automated driving. In *12th international conference on automotive user interfaces and interactive vehicular applications* (pp. 66–73).

Garate, V. R., Bours, R., & Kietlinski, K. (2012). Numerical modeling of ada system for vulnerable road users protection based on radar and vision sensing. In *2012 ieee intelligent vehicles symposium* (pp. 1150–1155).

Good, D. H., Krutilla, K., Chien, S., Li, L., & Chen, Y. (2017). Analysis of potential co-benefits for bicyclist crash imminent braking systems. In *Ieee international conference on intelligent transportation systems* (pp. 1–6).

Gu, W., Zhou, Z., Zhou, Y., Zou, H., Liu, Y., Spanos, C. J., & Zhang, L. (2017). Bikemate: Bike riding behavior monitoring with smartphones. In *Proceedings of the 14th eai international conference on mobile and ubiquitous systems: Computing, networking and services* (pp. 313–322).

Heinovski, J., Stratmann, L., Buse, D. S., Klingler, F., Franke, M., Oczko, M.-C. H., ... Dressler, F. (2019). Modeling cycling behavior to improve bicyclists' safety at intersections-a networking perspective. In *2019 ieee 20th international symposium on" a world of wireless, mobile and multimedia networks"(wowmom)* (pp. 1–8).

Hoermann, S., Kunz, F., Nuss, D., Renter, S., & Dietmayer, K. (2017). Entering crossroads with blind corners. a safe strategy for autonomous vehicles. In *2017 ieee intelligent vehicles symposium (iv)* (pp. 727–732).

Holländer, K., Schellenberg, L., Ou, C., & Butz, A. (2020). All fun and games: obtaining critical pedestrian behavior data from an online simulation. In *Extended abstracts of the 2020 chi conference on human factors in computing systems* (pp. 1–9).

Holländer, K., Wintersberger, P., & Butz, A. (2019). Overtrust in external cues of automated vehicles: an experimental investigation. In *Proceedings of the 11th international conference on automotive user interfaces and interactive vehicular applications* (pp. 211–221).

Jannat, M., Roldan, S. M., Balk, S. A., & Timpone, K. (2020). Assessing potential safety benefits of advanced pedestrian technologies through a pedestrian technology test bed. *Journal of Intelligent Transportation Systems*, *25*(2), 139–156.

Kiss, G., & Berecz, É. C. (2019). Questions of security in the world of autonomous vehicles. In *Proceedings of the 2019 the 5th international conference on e-society, e-learning and e-technologies* (pp. 109–115).

Kutela, B., Das, S., & Dadashova, B. (2022). Mining patterns of autonomous vehicle crashes involving vulnerable road users to understand the associated factors. *Accident Analysis & Prevention*, *165*, 106473.

Lin, A. Y., Kuehl, K., Schöning, J., & Hecht, B. (2017). Understanding" death by gps" a systematic study of catastrophic incidents associated with personal navigation technologies. In *Proceedings of the 2017 chi conference on human factors in computing systems* (pp. 1154–1166).

Liu, C., Fujishiro, R., Christopher, L., & Zheng, J. (2016). Vehicle–bicyclist dynamic position extracted from naturalistic driving videos. *IEEE transactions on intelligent transportation systems*, *18*(4), 734–742.

Mako, E. (2015). Evaluation of human behaviour at pedestrian crossings. In *2015 6th ieee international conference on cognitive infocommunications (coginfocom)* (pp. 443–447).

Moher, D., Liberati, A., Tetzlaff, J., Altman, D. G., & Group, P. (2009). Preferred reporting items for systematic reviews and meta-analyses: the prisma statement. *Annals of internal medicine*, *151*(4), 264–269.

Nassi, B., Mirsky, Y., Nassi, D., Ben-Netanel, R., Drokin, O., & Elovici, Y. (2020). Phantom of the adas: Securing advanced driver-assistance systems from split-second phantom attacks. In *Proceedings of the 2020 acm conference on computer and communications security* (pp. 293–308).

NHTSA. (2020). Overview of motor vehicle crashes in 2019. *Technical Report*.

Ponnaluri, R., & Alluri, P. (2021). Chapter 12 - examples of connected and automated vehicle applications. In (p. 181-203). Elsevier Inc.

Rettenmaier, M., Albers, D., & Bengler, K. (2020). After you?!–use of external human-machine interfaces in road bottleneck scenarios. *Transportation research part F: traffic psychology and behaviour*, *70*, 175–190.

Riaz, M. S., Cuenen, A., Janssens, D., Brijs, K., & Wets, G. (2019). Evaluation of a gamified e-learning platform to improve traffic safety among elementary school pupils in belgium. *Personal and Ubiquitous Computing*, *23*(5), 931–941.

Rostami, A., Cheng, B., Lu, H., Kenney, J. B., & Gruteser, M. (2016). Performance and channel load evaluation for contextual pedestrian-to-vehicle transmissions. In *Proceedings of the first acm international workshop on smart, autonomous, and connected vehicular systems and services* (pp. 22–29).

Sam, D., Evangelin, E., & Raj, V. C. (2015). A novel idea to improve pedestrian safety in black spots using a hybrid vanet of vehicular and body sensors. In *International confernce on innovation information in computing technologies* (pp. 1–6).

Sarker, A., & Shen, H. (2018). A data-driven misbehavior detection system for connected autonomous vehicles. *Proceedings of the ACM on Interactive, Mobile, Wearable and Ubiquitous Technologies*, *2*(4), 1–21.

Segata, M., Vijeikis, R., & Cigno, R. L. (2017). Communication-based collision avoidance between vulnerable road users and cars. In *2017 ieee conference on computer communications workshops* (pp. 565–570).

Sherony, R., & Zhang, C. (2015). Pedestrian and bicyclist crash scenarios in the us. In *2015 ieee 18th international conference on intelligent transportation systems* (pp. 1533–1538).

Sikkenk, M., & Terken, J. (2015). Rules of conduct for autonomous vehicles. In *Proceedings of the 7th international conference on automotive user interfaces and interactive vehicular applications* (pp. 19–22).

Sun, Z., Xing, Y., Wang, J., Gu, X., Lu, H., & Chen, Y. (2022). Exploring injury severity of vulnerable road user involved crashes across seasons: A hybrid method integrating random parameter logit model and bayesian network. *Safety science*, *150*, 105682.

Tahmasbi-Sarvestani, A., Mahjoub, H. N., Fallah, Y. P., Moradi-Pari, E., & Abuchaar, O. (2017). Implementation and evaluation of a cooperative vehicle-to-pedestrian safety application. *IEEE Intelligent Transportation Systems Magazine*, *9*(4), 62–75.

Utriainen, R. (2020). The potential impacts of automated vehicles on pedestrian safety in a four-season country. *Journal of Intelligent Transportation Systems*, *25*(2), 188–196.

Wang, M., Lundgren Lyckvi, S., & Chen, F. (2016). Why and how traffic safety cultures matter when designing advisory traffic information systems. In *Proceedings of the 2016 chi conference on human factors in computing systems* (pp. 2808–2818).

Wong, P. N. (2019). Who has the right of way, automated vehicles or drivers? multiple perspectives in safety, negotiation and trust. In *Proceedings of the 11th international conference on automotive user interfaces and interactive vehicular applications* (pp. 198–210).

Wotton, M. E., Bennett, J. M., Modesto, O., Challinor, K. L., & Prabhakaran, P. (2022). Attention all 'drivers': You could be to blame, no matter your behaviour or the level of vehicle automation. *Transportation research part F: traffic psychology and behaviour*, *87*, 219–235.